\documentstyle[12pt]{article}
\title{\bf Gauge theory of disclinations\\
on fluctuating elastic surfaces}
\author{ \vspace*{2mm}
{\bf E.A. Kochetov and V.A. Osipov} \\
\small \it Bogoliubov Theoretical Laboratory,\\
\small \it Joint Institute for Nuclear Research, 141980 Dubna, Russia}

\begin{document}
\date{}
\maketitle

\begin{abstract}

A variant of a gauge theory is formulated to describe disclinations on
Riemannian surfaces that may change both the Gaussian (intrinsic)
and mean (extrinsic) curvatures, which implies that both internal strains
and a location of the surface in $R^3$ may vary. Besides,
originally distributed disclinations are taken into account.
For the flat surface, an extended variant
of the Edelen-Kadi\'c gauge theory is obtained.
Within the linear scheme our model recovers
the von Karman equations for membranes, with a disclination-induced
source being generated by gauge fields.
For a single disclination on an arbitrary elastic surface a covariant
generalization of the von Karman equations is derived.

\end{abstract}
\vskip 1cm


\section{Introduction}

Elastic two-dimensional structures which are free to change their geometry
(membranes, thin films, etc.) as well as deformable materials with
spherical or tubular shapes (fullerenes, nanotubes)
are of considerable current interest (see,
e.g., \cite{nelson,chaikin,kuzmany}
and the references therein).
The properties of these crystalline structures are found to be essentially
affected by their topology. It has been found that topological defects,
first of all disclinations, play an important role in these objects.
In particular, the Kosterlitz--Thouless disclination unbinding
transition in hexatic membranes was shown to depend on shape
fluctuations~\cite{park1,deem} since a membrane
with a single disclination can lower its energy by buckling.
Fullerenes and nanotubes always contain at least 12 disclinations
on their surfaces (i.e. 5-fold coordinated sites) as a consequence
of the Euler theorem.
The effect of shape fluctuations
on the interaction of the disclinations on a spherical surface
with genus zero was studied in~\cite{park3}.

Elastic models for membranes and shells are well known
(see, e.g.,~\cite{landau,WS} and the references therein).
The main problem is how to incorporate
defects into the elastic theory of two-dimensional fluctuating surfaces.
A possible way has been considered in~\cite{nelson} where defects were
introduced {\it ad hoc} as source terms in the right-hand side of
the von Karman equations. This approach is efficient in description
of monolayers as well as membranes under condition that the bending
rigidity which controls out-of-plane fluctuations is small.
Similar approaches were developed in~\cite{park1} where
the Coulomb-like model was formulated. Disclinations
were introduced there as the point-like charges on the curved
surface.

The modern trends in theoretical description of topological defects
in condensed matter include geometrical and gauge-theory methods
(see, e.g., \cite{sadoc,holz,kleinert}).
In the present paper we put forward a gauge theory that enables one
to describe disclinations on $2D$ elastic surfaces that may
change both their intrinsic and extrinsic geometry. Namely, both
the internal strain and the location of the surface in $R^3$ may vary.
By analogy with the Edelen--Kadi\'c (EK) gauge model
of dislocation and disclinations~\cite{edelen,lagoud},
defects are incorporated
via dynamical gauge fields. However, we formulate
some principal statements which provide quite a new geometrical
setting to handle a problem of describing defect-developing deformations
of an elastic body. As is shown below,
this allows to describe defects
on arbitrarily fluctuating surfaces as well as dynamics of originally
distributed defects.
Notice that even in the $2D$ planar case our theory does not identically
coincides with that obtained within the EK approach~\cite{jpa93}.
As a matter of fact, it includes the latter as a particular case and
allows for a possibility to include objects with originally distributed
disclinations.

A basic motivation for our investigation has been that of extending the EK
gauge theory to include nontrivial geometry in the $2D$ case, e.g., that
of membranes, spheres, etc. It should be emphasized that
within the standard EK approach there is no room for such objects at all.
Indeed, within
that approach deformation is mathematically described as a diffeomorphic
mapping of a defect-free {\it domain} $D_0\in R^n$ which is usually called
a reference configuration to another one
$D\in R^n$ (current configuration) with noneucledian metric tensor $g$.
By the very definition $D_0$ and
$D$ necessarily have the same dimension $n$, thereby ruling out the
possibility to consider dynamics of, say, defects on $2D$-surfaces
for they are not domains in $R^3$.

Any attempt at generalizing the EK theory to study disclinations
on general manifolds should imply two steps. First, one has to appropriately
reformulate a classical theory of elasticity and, second, to introduce
in thus obtained new geometrical setting dynamical gauge fields.
A straightforward generalization might be that of considering a
diffeomorphic map $\chi:\,D_0\in\Sigma\to D\in\Sigma,$
where $\Sigma$ stands for a Riemannian surface \footnote{In particular,
$\Sigma$ may denote a {\it Riemann} surface, that is a $1D$ compact
orientable complex manifold.}
(i.e., a $2D$ real Riemannian manifold)
located in $R^3$.  This map can (locally) be described by functions
$\chi^a(x^b)\, (a,b=1,2)$, where $x^a$ denotes a point in $D_0$ and $\chi^a$
corresponds to coordinates in $D$.  A resulting Lagrangian to describe
elastic properties of the media would follow as a function of the state
vector $\chi^a(x)$, which would result in a kind of a field theory in the
nontrivial geometrical background. To include defects an appropriate gauge
field on a curved space $\Sigma$ would be further required, an obvious
complication in contrast with the EK approach.

The present approach has been proven to adequately incorporate
dynamics of disclinations on varying elastic surfaces as well
as to take into account originally distributed defects.
The key observation that considerably simplifies and at the same
time generalize the matter is that one should consider a
surfaces $\Sigma$ being embedded into a three-dimensional
flat space $R^3$ instead of proceeding entirely in terms of its intrinsic
geometry.  As we shall see shortly, following this idea will lead us to
a fairly plain and surprisingly complete theory of defect dynamics
on curved surfaces.

The paper is structured as follows. In section 2 we formulate the
gauge model of disclinations on $2D$ elastic surfaces. The action
which includes elastic deformations, self-energy of disclinations,
and the curvature energy is constructed in a self-consistent way.
A complete set of equations of motion is presented in section 3.
To illustrate the model, we consider three examples in section 4.
First, we derive equations of motion for the planar case and show
that they are distinct from those obtained within the EK approach.
Second, we study a problem of the fluctuating surface by employing the
linear approximation. In this case, the known equations for
fluctuating membranes are recovered in a self-consistent way
with a source formed by the gauge disclination fields.
Finally, a concrete realization of the model for a single disclination
on arbitrary elastic surface is presented.
Section 5 is devoted to concluding comments.

\section{The model}

Before proceeding, a few comments on the limitations of the theory as well
as on the conventions employed are to be made. First of all, we are solely
concerned with the $2D$ case, although our formalism can easily be extended
to any space dimensions. Our motivation is that the $2D$ case is both the
most important in applications and at the same time
the simplest one in notation.\footnote{$1D$ case is irrelevant
for us here since there are no $1D$ objects with an intrinsic curvature}
Secondly, we take into consideration only the rotational symmetry of
a system thereby making an attempt at describing disclinations and
leaving aside dislocations and other defects. By turning to the full-fledged
internal symmetry group instead of the orthogonal one the above restriction
might be avoided, though it will evidently entail considerable conventional
complications. Finally, static configurations are only considered, which
does not seem, however, to be a deficiency in the following exposition,
since including time evolution brings in no novel features compared
to the EK theory.

An action that is assumed to properly describe dynamics of disclinations
on a deformable elastic surface is taken in the form
\begin{equation}
S=S_{el}+S_{gauge}+S_{fl},
\label{eq:1.0}\end{equation}
where $S_{el}$ describes the elastic properties of the media, $S_{gauge}$
stands for the action of a gauge field that incorporates self-action
for disclinations, and $S_{fl}$ is the Helfrich-Canham
action~\cite{helf,canham} to describe the energy of a free fluctuating
surface.

Let us start by discussing the first piece of the action.
Let $x^a (a=1,2)$ be a set of local coordinates
on a certain Riemannian surface $\Sigma_0$.
(Indices $a,b,c,...=1,2$ are
tangent to $\Sigma_0$, whereas $i,j,k,...=1,2,3$
run over the basis of $R^3$). Under a deformation, $\Sigma_0$ is assumed
to evolve into some other surface $\Sigma$. To describe this we find
it convenient to introduce embeddings
$\Sigma_0,\,\Sigma\to R^3$ that can be realized in terms of two $R^3$-valued
functions $R_{(0)}^i(x^1,x^2)$ and $R^i(x^1,x^2)$, respectively. As the point
$(x^1,x^2)$ is varied vectors $\vec R_{(0)}$ and $\vec R$ sweep surfaces
$\Sigma_0$
and $\Sigma$, respectively. This is nothing but a familiar two-parametric
representation of surfaces in $R^3$, the point, however, being that
\begin{equation}
\vec R(x):=\phi^*\vec R_{(0)}=\vec R_{(0)}[\phi(x)],
\label{eq:1.1}\end{equation}
where $\phi^*$ is a pullback of $\phi:\,\Sigma_0\to\Sigma$.
In what follows functions $R^i_{(0)}(x^1,x^2)$ are chosen to
specify an initial configuration $\Sigma_0$, whereas dynamical
variables $R^i(x^1,x^2)$ are taken to describe the deformation $\Sigma_0\to
\Sigma$. With these conventions at hand, a proper
generalization of the elasticity theory turns out to be a straightforward
matter.

Representations for the induced metrics follow immediately
\begin{eqnarray}
g_{ab}&\equiv&(g_{\Sigma_0})_{ab}=
\partial_a\vec R_{(0)}\cdot
\partial_b\vec R_{(0)},\nonumber \\
\tilde g_{ab}&\equiv&(\phi^*g_{\Sigma})_{ab}
=(g_{\Sigma})_{cd}\frac{\partial{\phi^c}}{\partial x^a}\cdot
\frac{\partial{\phi^d}}{\partial x^b}=\frac{\partial{\vec R}}
{\partial{\phi^c}}\cdot\frac{\partial{\vec R}}{\partial{\phi^d}}\,
\frac{\partial{\phi^c}}{\partial x^a}\cdot
\frac{\partial{\phi^d}}{\partial x^b}=
\partial_a{\vec R}\cdot\partial_b{\vec R},
\label{eq:1.2}\end{eqnarray}
where the set $\{\phi^a\}$ stands for local coordinates on $\Sigma$.
The strain tensor is determined to be~\cite{kondo,bilby}
$$E_{ab}=\tilde g_{ab}-g_{ab}.$$
The elastic properties of the
deformed surface are described by an action
\begin{equation}
S_{el}= -\frac{1}{8}\int_{\Sigma_0} dx^1dx^2\sqrt g\left\{\lambda(tr E)^2
+2\mu\, trE^2\right\},
\label{eq:1.3}\end{equation}
where $tr E=g^{ab}E_{ab},\, g=det ||g_{ab}||$ and summation over repeated
indices is assumed. Let us mention that
we omit in (\ref{eq:1.3}) the terms of order $E^3$
and higher. In some special cases they can be considered as well
(see, e.g.,~\cite{lagoud}).

To proceed, a properly defined gauge field to describe disclinations
is to be introduced. To make this point transparent, let us step a bit
aside
and discuss the principle of a local gauge invariance from the geometric
viewpoint. In this regard, a simple example may be of some help.
Consider a scalar complex field $(\psi,\bar{\psi}):\,\,x\in R^n\to C$.
Let a Lagrangian exhibit a global $U(1)$ symmetry: $L\to L$ under
a transformation $\psi\to e^{i\alpha}\psi,\, \bar{\psi}\to e^{-i\alpha}
\bar{\psi}$. In what way the local gauge fields can then be introduced?
To this end, one considers $\psi$ and $\bar{\psi}$ as a description not of
a map $R^n\to C$ but of sections of a line $C$-bundle over $R^n$ with
the structure group $U(1)$. This is a trivial bundle which admits global
sections. A connection on this bundle
\footnote{To be more precise, a connection one-form is defined on the
associated principal bundle $P(R^n; U(1))$ and completely specifies the
covariant derivative on a $C$-bundle over $R^n$.}
is a familiar gauge
field $A_{\mu}$ which can be regarded as a $U(1)$ valued one-form.
The resulting theory is nothing but scalar electrodynamics.

In order to incorporate disclinations in the elasticity theory
(\ref{eq:1.3}), one should as the above example suggests consider the
$R^3$-vector bundle over $\Sigma$ and the $R^3_{(0)}$-bundle over
$\Sigma_0$ with the same structure groups $SO(3)$. The $so(3)$ valued one
form $ A^{(0)}_a(x)dx^a\,(A^{(0)}_a=\vec W^{(0)}_a\cdot\vec L, \quad
L^i\in so(3))$ serves as a connection one-form in the $R^3_{(0)}$-bundle
space over $\Sigma_0$, with $\vec W^{(0)}_a$ being the gauge potentials.
A connection on the $R^3$-bundle over $\Sigma_0$ is obtained by pulling
back the connection of the $R^3$-bundle over $\Sigma$:  $$\vec W_a:=
\phi^*\,\vec W_a\mid_{\Sigma}=\partial_a\phi^b(\vec W_b\mid_{\Sigma}).$$ By
replacing in (\ref{eq:1.3}) ordinary derivatives $\partial_a\vec R$ and
$\partial_a\vec R_{(0)}$ by the covariant ones $\nabla_a\vec R=
\partial_a\vec R + [\vec W_a,\vec R]$ and $\nabla_a\vec
R_{(0)}=\partial_a\vec R_{(0)} + [\vec W^{(0)}_a,\vec R_{(0)}]$,
respectively, one arrives at the desired locally $SO(3)$ invariant
representation for the elasticity Lagrangian.

A few remarks are in order at this stage. First, we consider
potentials $\vec W^{(0)}_a$ as given fixed functions
to describe disclinations originally distributed on $\Sigma_0$. These
potentials being involved in a full set of equations of motion,
provide a possibility to keep track of the dynamics of these disclinations
under deformation.
Such a possibility is missing in the standard EK theory where from the
very beginning only defect-free initial configurations are allowed.

Second, although essentially two-dimensional manifolds are considered,
{\it three}-dimensional rotational group $SO(3)$
is involved, which seems to be quite natural in the framework of our
approach. Enlarging the structure group to the semidirect product
$SO(3)\rhd T(3)$, where $T(3)$ stands for the group of translations
in $R^3$, enables one to include into consideration both disclinations
and dislocations.

Third, if we made an attempt to formulate an appropriate gauge theory in the
scope of the direct generalization of the EK approach, we would encounter
the following problem. With the $\chi^a(x)$ being not the $\Sigma$-valued
functions defined on $\Sigma$ but local sections of a $\Sigma$-bundle
over $\Sigma$, the derivatives $\partial_a\chi_b$ must be replaced by
suitable covariant derivatives. What would be the "gauge group" in the
definition of these covariant derivatives? In general this would be the
infinite dimensional group Diff~$\Sigma$ of all diffeomorphisms of $\Sigma$
which is a fibre of the bundle.
For an arbitrary $\Sigma$ it almost surely does not
include rotations in contrast to the group Diff~$R^3=GL(3,R) \ni SO(3)$.
To find a way out, one might be tempted to relate defects
to a group of all isometries of $(\Sigma,g)$ rather than to $SO(3)$. This
group belongs to Diff $\Sigma$ and is generated by operators ${\hat
V}_a$ obeying $${\cal {L}}_{V_a}\,g=0,$$ where ${\cal {L}}_{V_a}$ denote
Lie derivatives along corresponding vector fields $V_a$.  Needless to say
that these equations cannot be in general solved explicitly, except in a
few simple instances, e.g., planes, spheres and other highly symmetrical
objects. In a general case, however, and as long as the $SO(3)$ group
is taken to be relevant for describing disclinations, this approach seems
to pose a problem.

Let us now turn to the second piece of the whole action, $S_{gauge}$, which
describes a self-energy of disclinations. It acquires a standard form
of the $SO(3)$ Yang-Mills action:
\begin{equation}
S_{gauge}=-\frac{s}{4}\int_{\Sigma_0}\,\sqrt g dx^1dx^2\langle
{\cal F}^{ab},{\cal F}_{ab}\rangle,
\label{eq:1.4}\end{equation}
where $s$ is a coupling strength, the form
$\langle,\rangle$ stands for the $so(3)$ Killing trace and the
$so(3)$-valued curvature tensor
${\cal F}_{ab}=\vec F_{ab}\cdot\vec L,\quad
\vec F_{ab}=\partial_a\vec W_b-\partial_b\vec W_a
+[\vec W_a,\vec W_b]$. The $so(3)$ generators $L_i$
obey the following commutation rules
$$ [L_i,L_j]=\epsilon_{ij}^k\,L_k,\quad  i,j,k=1,2,3,$$ where
$\epsilon_{ij}^k$ stands for the fully antisymmetric tensor in $R^3$.

Finally, the Helfrich-Canham action that describes a self-energy of a
fluctuating two-surface in $R^3$ looks like~\cite{helf,canham}
\begin{equation}
S_{fl}=\frac{\kappa}{2}\int_{\Sigma_0}\,\sqrt g dx^1dx^2 \,(tr K)^2
+ \frac{\kappa_G}{2}\int_{\Sigma_0}\sqrt g\, dx^1dx^2 \, det\,g^{ab}K_{bc}
\label{eq:1.5}\end{equation}
where $\kappa$ is a bare bending rigidity and $\kappa_G$ is a Gaussian
rigidity,
\begin{equation}
K_{ab}=\vec N\cdot D_aD_b\vec R \label{eq:1.55}\end{equation} is
the curvature tensor, and $\vec N$ is the unit normal to the
surface $$\vec N=\frac{[\partial_1\vec R,\partial_2\vec R]}
{|[\partial_1\vec R,\partial_2\vec R]|}.$$ The covariant
derivative $$D_a:=\partial_a+\Gamma_a$$ includes the Levi-Civita
connection $\Gamma_a$ to be written down explicitly in the next
section. Two scalar functions enter (\ref{eq:1.5}): $tr
K=g^{ab}K_{ab}$ called the mean (extrinsic) curvature, and
$S=det\,g^{ab}K_{bc}$ referred to as the Gaussian (intrinsic)
curvature. In view of the fact that the second piece on the r.h.s.
of (\ref{eq:1.5}) is a topological invariant and depends only on
the genus of the surface, it does not affect classical equations
of motion. Incorporating disclinations amounts then to performing
in the above formulas a substitution $$\partial_a\to\nabla_a,$$
which results in the observation that the both pieces of action
(\ref{eq:1.5}) will now contribute to the equations of motion.

\section{Equations of motion}

Equations of motion follow from the Hamilton principle of stationary action
$\delta S=0$ and read
\begin{eqnarray}
{\cal D}_b\,\vec{\sigma}^b+\vec J=0,
\label{eq:2.1}\end{eqnarray}
\begin{equation}
s{\cal D}_a\,\vec F^{ab}-\frac{1}{2}\,[\vec R,\vec{\sigma}^b]+
\frac{1}{2}\,\vec I^b=0
\label{eq:2.2}\end{equation}
where a set of the stress vectors\footnote{As was already mentioned,
index $a$ corresponds to the tangent space $T\Sigma_0$,
whereas $i$ is referred to the basis of the underlying space
$R^3$. These are different manifolds and consequently $\vec\sigma^a=
\{\sigma^a_i\}$ are viewed as a {\it set of vectors} in $R^3$. On the
other hand, these spaces coincide in the EK approach
and $\sigma^a_i$ reduces to a stress {\it tensor}.}
\begin{equation} \vec{\sigma}^b=\frac{1}{2}(\nabla_a\vec
R)\,\rho^{ab}, \label{eq:2.3}\end{equation} with
\begin{equation}
\rho^{ab}:=\lambda\,g^{ab}\,tr E +2\mu\,E^{ab}
\label{eq:2.21}\end{equation}
has been
introduced.  The total covariant derivative $${\cal D}_a:=\nabla_a+\Gamma_a$$
includes the Levi-Civita (torsion-free, metric compatible) connection
\begin{equation}
\Gamma^b_{ac}:=(\Gamma_a)^b_c = \frac{1}{2}g^{bd}
\left(\frac{\partial g_{dc}}{\partial x^a}+\frac{\partial g_{ad}}
{\partial x^c}-\frac{\partial g_{ac}}{\partial x^d}\right)
\label{eq:2.3a}\end{equation}
to take care of a metric factor $\sqrt g$ in Eqs.
(\ref{eq:1.3}-\ref{eq:1.5}) when a variation of the total action is
calculated. It is noteworthy that $\Gamma_a$ depends on $\vec W^{(0)}_a$
that enters in a definition of $g_{ab}$:
\begin{eqnarray}
g_{ab}&=&\nabla_a\vec R^{(0)}\cdot\nabla_b\vec R^{(0)}
=\partial_a\vec R_{(0)}\cdot
\partial_b\vec R_{(0)}+
\partial_a\vec R^{(0)}[\vec W^{(0)}_b,\vec R^{(0)}] +
\partial_b\vec R^{(0)}[\vec W^{(0)}_a,\vec R^{(0)}] \nonumber \\
&+&(\vec W^{(0)}_a
\vec W^{(0)}_b)\vec
R_{(0)}^2 - (\vec W^{(0)}_a\vec R_{(0)})(\vec W^{(0)}_b\vec R_{(0)}).
\label{eq:2.4}\end{eqnarray}
We have also abbreviated
$$\vec J=\frac{1}{\sqrt g}\frac{\delta S_{fl}}{\delta\vec R},\quad
{\vec I}^b=\frac{1}{\sqrt g}\frac{\delta S_{fl}}{\delta{\vec W}_b},$$
bearing in mind that
explicit formulae are available in particular cases.

Obviously, for the elastic plane without defects (\ref{eq:2.1})
reduces to the well-known equilibrium equation $\partial_b\vec\sigma^b=0$
while (\ref{eq:2.2}) is absent. In the general case,
both the gauge fields and the affine connection enter
(\ref{eq:2.1}) and (\ref{eq:2.2}) thus affecting stress fields.
In view of (\ref{eq:2.3a})
$$ \Gamma^{b}_{ba}=\frac{1}{2}\frac{\partial}{\partial x^a}\log g,\quad
g=det ||g_{ab}||$$ and one may consequently rewrite
Eqs.~(\ref{eq:2.1},\ref{eq:2.2}) in the form
\begin{equation}
\frac{1}{\sqrt g}\partial_b\,(\sqrt g\vec{\sigma}^b)+[\vec W_b,
\vec{\sigma}^b]+\vec J =0
\label{eq:2.1a}\end{equation} and
\begin{equation}
\frac{1}{\sqrt g}\partial_a\,(\sqrt g
F^{ab})+\frac{1}{2s}[\vec{\sigma}^b,\vec R]+\frac{1}{2s}\,\vec I^b=0,
\label{eq:2.1b}\end{equation} respectively.
As is seen, the basic self-consistent equations of the model
are strongly nonlinear and it is difficult to get the general
solution of the problem. It will be shown below,
however, that for some physically interesting problems
these equations become essentially simpler and can be
solved explicitly.

\section{Applications}

In this section we intend to consider three explicit realizations
of the above theory. First, we show how the standard $2D$ EK theory
emerges in the framework of our approach. Second, we
consider an important in applications case of fluctuating elastic
membranes, and, finally, a single disclination on arbitrary elastic
surface is examined.

\subsection{\it Disclinations on elastic plane}

Let us start by recalling an explicit
formulation of the $2D$ EK theory~\cite{jpa93}.
Consider a diffeomorphic map $\chi:\,D_0\to D,$ where $D_0\in R^2$ appears
as a strain free (which means $g_{ab}=\delta_{ab}$)
domain and $D\in R^2$ stands for its image under deformation.
Evidently, $\chi^a$ can be viewed upon as $R^2$-vector valued functions.
Resulting action that describes dynamics of disclinations on a planar
elastic body consists of two pieces $S_{el}$ and $S_{gauge}$ given by Eqs.
(\ref{eq:1.3}) and (\ref{eq:1.4}), respectively, provided one puts
$$E_{ab}=\vec B_a\cdot\vec B_b-\delta_{ab},\quad
\vec B_a=\partial_a\vec{\chi}+[\vec W_a,\vec{\chi}],$$ where
$\vec W_a=(0,0,W_a)$. A complete set of the Euler-Lagrange equations
can be easily derived and shown to possess exact vortex-like
solutions~\cite{pla92}.

On the other hand, basic notation of our theory in this case looks
as follows. First we have the embeddings:
$\vec R_{(0)}=(x,y,0)$ and
$\vec R=(R_1(x,y),R_2(x,y),0)$, where $x,y$ are Cortesian
coordinates. Evidently, one has
$\vec W_a=(0,0,W_a),\, \vec F^{ab}=(0,0,F^{ab})$, where
$F_{ab}=\partial_aW_b-\partial_bW_a$. The same representation holds
for $\vec W_a^{(0)}$ which appears as a fixed gauge field
associated with originally distributed disclinations.

Equations of motion (\ref{eq:2.1}) and ({\ref{eq:2.2}) take the form
$(\vec J=\vec I^a=0)$
$$\frac{1}{\sqrt g}\partial_b\,(\sqrt g\vec{\sigma}^b)+[\vec W_b,
\vec{\sigma}^b]=0$$
$$\frac{1}{\sqrt g}\partial_a\,(\sqrt g
F^{ab})+\frac{1}{2s}[\vec{\sigma}^b,\vec R]=0,$$
where $g=det ||g_{ab}||$ and in view of Eq. (\ref{eq:2.4})
$$g_{ab}=\delta_{ab}+
\epsilon_{\alpha a}W^{(0)}_bR_{(0)}^{\alpha}+
\epsilon_{\beta b}W^{(0)}_aR_{(0)}^{\beta} + (W^{(0)}_aW^{(0)}_b)\vec
R_{(0)}^2, \quad \alpha,\beta=1,2.$$ Metric tensor $g_{ab}$ is seen to
deviate from its flat counterpart $\delta_{ab}$ at $W^{(0)}\neq 0$,
whereby a nontrivial geometry is dynamically generated.  It is at this
point that our theory deviates from the standard EK one even in the
trivial planar case.

At $W^{(0)}=0$ the above equations reduce to
those of the $2D$ EK theory~\cite{jpa93}.
In the linear approximation~\cite{pla94} no interaction between
$W_a$ and $W^{(0)}_a$ fields occurs, so that a total disclination
flow $\oint\vec{\cal W}d\vec l$ that determines a source which gives
rise to disclination-induced displacements, is found to be
$\vec{\cal W}= \vec W-\vec W^{(0)}.$

\subsection{\it Disclinations on membranes}

Let us now turn to the gauge theory of defects for fluctuating membranes
starting from
(\ref{eq:2.1},\ref{eq:2.2}),
our aim being first that to recover a conventional Landau theory
without defects.
No disclinations are assumed to be originally distributed as well,
so that we put
$\vec W_a^{(0)}=0$. A flat membrane that fluctuates in the $z$-direction
can be described by the embeddings:
$$\vec R^{(0)}=(x,y,0),\quad \vec R=(x+u_x, y+ u_y,f),$$ where
$\vec R=\vec R^{(0)} +\vec U$ and $\vec U=(u_x,u_y,f)$ is a displacement
of the $(x,y,0)$ point under deformation. (In writing out the $\vec U$
components we stick to the Landau notation~\cite{landau}.)
In accordance with the Landau
theory $\vec U$ is assumed to be small in the transverse directions,
along with the requirement that
its $z$-component, $f(x,y)$, is to be viewed as a slowly varying function.
Precise restrictions on components of $\vec U$ directly follow
those in~\cite{landau}. It is clear by obvious reasoning that
within the adopted approximations only
the $z$-component of $\vec W_a$ matters, so that we may
put $\vec W_a=(0,0,W_a)$. The strain tensor is then determined to be
\begin{eqnarray}
E_{ab}&=&\nabla_a\vec R\cdot\nabla_b\vec R-\delta_{ab}\nonumber \\
&=& \partial_au_b+\partial_bu_a +\partial_af\partial_bf+
\epsilon_{\alpha a}W_bR_{(0)}^{\alpha} +
\epsilon_{\alpha b}W_aR_{(0)}^{\alpha} +{\cal O}(u^2,u\partial f, W^2).
\label{eq:3.1}\end{eqnarray}
with Greek indices $\alpha,\beta=1,2$ being used to specify
coordinates of the plane orthogonal to the $z$-axes.

As for the curvature tensor $K_{ab}$, it is easily calculated to be
\begin{equation}
K_{ab}= \partial_af\partial_bf+{\cal O}(u^2,u\partial f, W^2).
\label{eq:3.2}\end{equation}
Within this accuracy one, consequently, obtains
$$ tr K=\Delta f.$$

To proceed with the equations of motion, we write down Eq.~(\ref{eq:2.1})
in components referred to the
$z$-direction and transverse plane, respectively:
$$\nabla_b{\sigma}^b_3+\kappa J_3=0$$
and
$$\nabla_b{\sigma}^b_{\alpha}+\kappa J_{\alpha}=0.$$
Vector $\vec J$ is easily found to be $\vec J=(0,0,\Delta^2f)$, which yields
for the above equations
\begin{equation}
\frac{1}{2}\partial_b(\partial_af\,\rho^{ab})+\kappa\,\Delta^2f=0,
\label{eq:3.3}\end{equation}
and
\begin{equation}
\partial_b\rho^{b\beta}=0,
\label{eq:3.4}\end{equation}
respectively.
Equation (\ref{eq:2.21}) now reads
$$\rho^{ab}=\lambda\,tr E\, \delta^{ab} +2\mu\,E^{ab},$$
with $E^{ab}$ being given by (\ref{eq:3.1}).

Within the present approximation, $\vec W_a=(0,0,W_a),$
and Eq.~(\ref{eq:2.2}) is found to go over into
\begin{equation}
\partial_aF^{ab}=\frac{1}{4s}\epsilon_{\alpha\beta}\rho^{\beta b}
R_{(0)}^{\alpha}.
\label{eq:3.5}\end{equation}
In deriving this, we have also used that $I^a_3=0$.
Equations (\ref{eq:3.3},\ref{eq:3.4}) coincide exactly at $W_a=0$
with those in~\cite{landau}. In our case, however, there appears an
additional equation (\ref{eq:3.5}) which ensures the
self-consistency of the model. As will be shown below,
in the linear approximation this equation includes only
gauge fields. In this case, one can use its solutions to
determine the disclination-induced sources for the remaining equations.

Notice that the foregoing procedure which allowed to obtain
(\ref{eq:3.3} -- \ref{eq:3.5}) is in agreement with the
linearization scheme proposed within the EK model~\cite{edelen}.
This scheme is based on a homogeneous scaling
of the gauge group generators (see also details in~\cite{lagoud,jpa91}).
It can be shown that by applying this procedure to
(\ref{eq:2.1},\ref{eq:2.2}) one can get (\ref{eq:3.3}--\ref{eq:3.5}).
It is important to note, however, that
in accordance with this procedure one has to choose properly
the relation between the model parameters ($\lambda/s$ and $\mu/s$).
Depending on this choice one can describe elastic media with
different properties. The classical elasticity theory which
is of our interest here is recovered
in the limit $\lambda/s\sim\epsilon$ and $\mu/s\sim\epsilon$ where
$\epsilon$ is a scaling parameter.
In this case, (\ref{eq:3.5}}) reduces to
\begin{equation}
\partial_aF^{ab} = 0.
\label{eq:3.6}\end{equation}
A singular vortex-like solution of (\ref{eq:3.6}) reads~\cite{pla92}
\begin{equation}
W_b = -\nu\epsilon_{bc}\partial_c\log r,
\label{eq:3.7}\end{equation}
where $\nu$ is the Frank index.

One can easily see that (\ref{eq:3.3}) is rewritten as
\begin{equation}
(\partial_b\partial_a f)\sigma^{ab} = -\kappa^2\Delta^2f,
\label{eq:3.8}\end{equation}
where (\ref{eq:3.4}) is taken into account. Let us
consider (\ref{eq:3.4}). It resembles the usual equilibrium
condition but the stress tensor includes the gauge fields.
Within the linear approximation terms with $W_b$ can be separated
thus forming the source in the right-hand side of (\ref{eq:3.4}).
In particular, for the planar case one can reproduce the known
exact solution for a straight wedge disclination~\cite{pla94}.

To compare our results with those in~\cite{nelson} let us
differentiate Eq. (\ref{eq:3.4}), which yields
$$\partial_{\beta}\partial_b\rho^{b\beta}=0.$$
After straightforward calculations one can rewrite this equation as
\begin{equation}
(\lambda/4\mu+1/2)\,\Delta\,tr E = (\partial_x\partial_y f)^2-
\partial_x^2 f\partial_y^2 f+
\epsilon_{\alpha b}\partial_\alpha W_b.
\label{eq:3.9}\end{equation}
Notice that the last term in the right-hand side of (\ref{eq:3.9})
describes a source due to disclination fields. For solution
(\ref{eq:3.7}) it takes the form
$$\epsilon_{\alpha b}\partial_\alpha W_b=
\nu\Delta\log r=2\pi\nu\delta (\vec r).$$
Introducing the Airy stress function $\chi(\vec r)$ by
$\sigma_{b\alpha}=:\epsilon_{bm}\epsilon_{\alpha n}\partial_m
\partial_n\chi(\vec r)$,
one can finally rewrite equations (\ref{eq:3.8}) and (\ref{eq:3.9}) as
\begin{eqnarray}
\kappa\Delta^2f = (\partial_y^2\chi)(\partial_x^2 f)+(\partial_x^2\chi)
(\partial_y^2 f)-
2(\partial_x\partial_y\chi)(\partial_x\partial_y f), \nonumber \\[0.2cm]
K_0^{-1}\Delta^2\chi = (\partial_x\partial_y f)^2-(\partial_x^2 f)
(\partial_y^2 f)+ 2\pi\nu\delta (\vec r),
\label{eq:3.10}\end{eqnarray}
respectively. Here $K_0=4\mu(\lambda +\mu)/(\lambda +2\mu)$, and
$trE=(1/(\lambda+\mu))\Delta\chi(\vec r)$.
As is seen, (\ref{eq:3.10}) are exactly the von Karman equations given
in~\cite{nelson} for defects in hexatic membranes.
It should be mentioned once more that the source term in (\ref{eq:3.10})
is not appeared {\it ad hoc} but
generated by the gauge fields due to a disclination. Its
exact form follows from the self-consistent solution of the
basic model equations.
An analysis of (\ref{eq:3.10}) shows~\cite{nelson}
that isolated positive
(five-fold) disclinations on free membranes buckle into a cone,
while the negative (seven-fold) disclination leads to a saddle surface.
The energy of a positive disclination was found to be less than
that of a negative one. It is interesting that this
asymmetry is absent in flat membranes and monolayers.
Notice that the linear approximation used
in this section allows to properly describe only the small
out-of-plane fluctuations. Otherwise, the full set of
equations (\ref{eq:2.1},\ref{eq:2.2}) should be examined.

\subsection{\it Single disclination on arbitrary elastic surface}

In this subsection we demonstrate a non-trivial realization
of the proposed model: a case of a single disclination on elastic
surface. Let us
consider a surface $\Sigma_0$ in a three-dimensional Euclidean space $R^3$.
For any point $p\in\Sigma_0$ choose the $z$-axis to be normal to
a tangent plane at $p$. Having in that way fixed the coordinate system
in $R^3$, we consider in what follows $\Sigma_0$ as an embedding
$$(u,v)\to (R_{(0)}^x(u,v), R_{(0)}^y(u,v), R_{(0)}^z(u,v)),$$
with $x^1=u, x^2=v$ being the local coordinates on $\Sigma_0$.
Under a local deformation concentrated at $p$, any nearby point undergoes
a displacement
$$\vec R_{(0)}(u,v)\to\vec R(u,v) =(\vec R_{(0)}^{\bot}(u,v)+
\vec U^{\bot}(u,v),\,
R^z(u,v)).$$  As is above, the transverse displacement $\vec U^{\bot}$
is assumed to
be small compared to the vertical one $U^z:= R^z-R_{(0)}^z$, the latter
being considered as a
slowly varying function on a local chart that contains $p$.
One may also put $\vec W_a=(0,0,W_a)$, which is in agreement with the
assumptions made. The strain tensor then becomes
\begin{equation}
E_{ab}=\partial_a\vec R^{\bot}_{(0)}\partial_b\vec U^{\bot}
 +\epsilon_{\alpha\beta}
\partial_aR^{\beta}_{(0)}R^{\alpha}_{(0)} W_b +(a\leftrightarrow b)
+\partial_aR^z\partial_bR^z-\partial_aR_{(0)}^z\partial_bR_{(0)}^z
+{\cal O}(U_{\bot}^2,W^2, U_{\bot}W).
\label{eq:4.1}\end{equation}

In order to make a more close connection with the gauge theory of defects
for
fluctuating membranes of the preceding section, we find it appropriate
to consider here a coordinate $R^z$ as a dynamical variable rather than
its displacement, $U^z$. In particular, the curvature tensor
$K_{ab}$ appears then as a direct covariant generalization of
Eq. (\ref{eq:3.2}). Indeed, it is clear that
$$\vec N^{(0)}_p:={\frac{[\partial_u\vec R_{(0)},\partial_v\vec R_{(0)}]}
{|[\partial_u\vec R_{(0)},\partial_v\vec R_{(0)}]|}}_{\mid_p}=(0,0,1).$$
On the other hand, we have
$${K_{ab}}_{\mid_{p'}}=\vec N_{p'}\cdot D_aD_b\vec R_{\mid_{p'}}=
(\vec N_p+\delta\vec N_p)\cdot D_aD_b\vec R_{\mid_{p'}}$$
$$=\vec N_p\cdot D_aD_b\vec R_{\mid_{p'}}
+{\cal O}\left(||\delta\vec N_p||\right),
\quad \delta\vec N_p:=\vec N_{p'}-\vec N_p.$$
As $\vec N_p$ are sufficiently smooth and slowly varying functions of a
reference point $p$, there exists a certain neighborhood of $p$, $V_p$,
such that
$||\delta\vec N_p||\ll 1$ for any
$p'\in V_p$.

In the linear approximation $\vec N_p$ can be replaced by its unperturbed
value $\vec N^{(0)}_p$, so that
for a local deformation of $\Sigma_0$ in the vicinity
of $p$
one obtains
$$K_{ab}=D_aD_bR^z$$ and consequently
$$tr\,K=g^{ab}D_aD_b R^z=D_aD^aR^z=:\Delta_{cov}R^z,$$
where the covariant Laplacian operator has been introduced.

Equation of motion (\ref{eq:2.1}) in the $z$ and transverse
components now reads
\begin{equation}
\frac{1}{2}\,D_b\left[(\partial_aR^z)\rho^{ab}\right]+
\kappa\,\Delta^2_{cov}R^z=0,
\label{eq:4.2}\end{equation}
and
\begin{equation}
D_b\left[(\partial_a\vec R^{\bot}_{(0)})\rho^{ab}\right] =0,
\label{eq:4.3}\end{equation}
respectively\footnote{Components of the three-vectors $\vec R(u,v)$
and $\vec R_{(0)}(u,v)$
appear as {\it scalar} functions with respect to a change of coordinates
on $\Sigma_0$ and hence there is no need to
replace the ordinary derivatives by the covariant ones in Eqs.~(\ref{eq:4.2},
\ref{eq:4.3}).}.
In the Monge gauge, $\vec R_{(0)} = (R^x_{(0)}=u=:x, R^y_{(0)}=v=:y,
R^z_{(0)}(x,y))$,
the last equation is written as
\begin{equation}
D_b\rho^{ab} = 0.
\label{eq:4.4}\end{equation}
As is seen, (\ref{eq:4.2}) and (\ref{eq:4.4}) are nothing
but a formally covariant representation
of the Landau equations (\ref{eq:3.3}) and (\ref{eq:3.4})
with the nontrivial geometry of $\Sigma_0$ taken into account.
These equations should be accompanied by the field equation (\ref{eq:2.2})
which in the linear approximation describes the $SO(2)$ gauge field in a
curved background:
\begin{equation}
D_aF^{ab}=0,\quad F^{ab}=\partial^aW^b-\partial^bW^a.
\label{eq:4.5}\end{equation}
The following steps are just the same as in the previous subsection.
A singular solution of (\ref{eq:4.5})
takes the form
\begin{equation}
W^b = -\nu\varepsilon^{bc}D_c G(x,y),
\label{eq:4.6}\end{equation}
where $G(x,y)$ satisfies the equation $$D_aD^aG(x,y) =
2\pi \delta^2(x,y)/\sqrt g,$$
with $\varepsilon_{ab}=\sqrt{g}\epsilon_{ab}$ being
the fully antisymmetric tensor on $\Sigma_0$. Note also that a
conventional "flat" $\delta$-function $\delta^2(x,y):=\delta(x)\delta(y)$
is to be multiplied by a metric factor $1/\sqrt g$ to make a scalar
under a coordinate reparametrization.

Equation (\ref{eq:4.2}) now reads
\begin{equation}
\frac{1}{2}(D_bD_a R^z)\rho^{ab} = -\kappa^2\Delta^2_{cov}R^z,
\label{eq:4.7}\end{equation}
while (\ref{eq:4.4}) can be differentiated
$D_aD_b\rho^{ab}=0$ to yield
\begin{equation}
(\lambda/2\mu+1)\,\Delta_{cov}\,tr E =
(D_aD_b R^z)^2 - (\Delta_{cov}R^z)^2 - (R^z \leftrightarrow
R^z_{(0)}) + 4\pi\nu\delta^2(x,y)/\sqrt g,
\label{eq:4.8}\end{equation}
where the condition $\varepsilon^{ab}D_aW_b=\nu\Delta_{cov}G(x,y)$ has been
taken into account.
Obviously, (\ref{eq:4.8}) is the covariant analog of (\ref{eq:3.9}).
It should be mentioned that
due to the specific choice of the coordinate system related to
a certain reference point, Eqs.~(\ref{eq:4.7}) and (\ref{eq:4.8})
describe a {\it single} defect located at this point. In this regard,
adding at least one more disclination makes
the situation more difficult, so that turning to the basic equations
of motion (\ref{eq:2.1}) and (\ref{eq:2.2}) seems necessary.

\section{Conclusion}

The model developed in this paper allows to describe
disclinations on arbitrary elastic surfaces. It includes
Riemannian surfaces that may change their geometry under
deformations.
In particular, within the proposed model one can study elastic properties
of various materials containing disclinations:
monolayers (flat surface), membranes (curved surface),
fullerenes (spherical surface) as well as nanotubes
(which can be considered as deformed spheres).
For the flat surface, we have obtained the extended variant
of the EK gauge theory that includes originally distributed disclinations.
Within the linear scheme our model recovers
the von Karman equations for membranes with a disclination-induced
source being generated by gauge fields.
For a single disclination on an arbitrary surface a covariant
generalization of these equations is obtained.

In our opinion, the most intriguing application of this theory might be
that for a disclination on a sphere. The obtained equations
(\ref{eq:4.2}) and (\ref{eq:4.4}) are the most general ones
which allow to study this problem properly. Notice, however, that
this is a challenging task. Indeed, an analytical solution
of the simpler system (\ref{eq:3.3}) and (\ref{eq:3.4}) has been obtained
only in the limiting case $K_0\rightarrow\infty$. Nevertheless, there are
precise numerical solutions of (\ref{eq:3.3}) and (\ref{eq:3.4})
for arbitrary $K_0$ (see, e.g.,~\cite{nelson} and the references therein).
Obviously, any attempts to solve (\ref{eq:4.2}) and (\ref{eq:4.4}),
either analytically or numerically would be of great interest.

Another important problem relevant for the physics of fullerenes
and carbon nanotubes concerns the electronic properties of
these materials. An attempt at extending the EK gauge theory of defects
to include fermionic fields has been made in~\cite{pa91}.
As has been shown, the self-consistent gauge model allows to describe
physically interesting effects: the Aharonov-Bohm-like
electron scattering due to disclinations~\cite{pla92},
an electron localization
near topological defects~\cite{jpa91,jpc95} as well as a formation of the
polaron-type states near dislocations~\cite{prb95}.
We expect that incorporating fermions in the above-formulated
theory may provide a new insight into disclination theory
in a curved background in the presence of electrons and reveal
some novel physical phenomena. Fortunately, our approach
allows for a natural extension of the model to include fermions.
Indeed, the model action (\ref{eq:1.0}) is constructed out of
sections of the vector bundles over $\Sigma_0$.
Considering on the other hand a spin bundle over $\Sigma_0$, that is a
tangent bundle $T\Sigma_0\to\Sigma_0$ with a structure group $Spin(2)$
amounts to incorporating fermions into the theory, with fermionic fields
being local sections of the spin bundle. (We assume that $\Sigma_0$ admits
a spin structure, which is the case, for instance, if $\Sigma_0$ appears as
a Riemann surface with genus $g$.) Since $dim \Sigma_0=2$, the spin
connection
term drops out from an action, so that the Dirac operator on $\Sigma_0$
takes the form
$$D=i\gamma^a\partial_a +m, \quad \{\gamma^a,\gamma^b\}=g^{ab},$$
where $\gamma^a=e^a_{\alpha}\gamma^{\alpha},\,
g_{ab}=e^{\alpha}_a e^{\beta}_b\,\delta_{\alpha\beta}$ and the spin group
$Spin(2)$ is generated by two Dirac matrices $\gamma_{\alpha},\,\alpha=1,2$
which can be taken as the Pauli matrices $\sigma_1$ and $\sigma_2$.
An explicit form for the fermion action as well as a full set of ensuing
equations of motion will be given elsewhere.

\vskip 0.5cm
This work has been supported by the Russian Foundation
for Basic Research under grant No. 97-02-16623.



\begin{thebibliography}{99}



\bibitem{nelson} Nelson D R 1994 {\it Defects in Superfluids, Superconductors,
         and Membranes (NATO ASI on Fluctuating Geometries
                 in Statistical Physics and Field Theory)} (Les Houches)
\bibitem{chaikin} Chaikin P and Lubensky T C 1995 {\it Principles
                  of Condensed Matter Physics} (Cambridge: Cambridge
                  University Press)
\bibitem{kuzmany} Kuzmany H, Fink J, Mehring M and Roth S (ed) 1996
                  {\it Fullerenes and Fullerene Nanostructures}
          (Singapore: World Scientific)
\bibitem{park1} Park J M and Lubensky T C 1996 {\it Phys. Rev.} E {\bf 53}
           2648; ibid {\bf 53} 2665
\bibitem{deem} Deem M W and Nelson D R 1996 {\it Phys. Rev.} E
                 {\bf 53} 2551
\bibitem{park3} Park J M 1996 {\it Phys. Rev.} E {\bf 54} 5414
\bibitem{landau} Landau L D and Lifshitz E M 1960 {\it Theory of Elasticity}
         (Oxford: Pergamon)
\bibitem{WS} Nelson D R, Piran T and Weinberg S (ed) 1989
                {\it Statistical Mechanics of Membranes and Surfaces}
                (Singapore: World Scientific)
\bibitem{helf} Helfrich W 1973 {\it Z. Naturforsch.} C {\bf 28} 693
\bibitem{canham} Canham P 1970 {\it J. Teo. Bio.} {\bf 26} 61
\bibitem{sadoc} Sadoc J F (ed) 1990 {\it Geometry in Condensed Matter
                Physics} (Singapore: World Scientific)
\bibitem{holz}  Holz A 1992 {\it J. Phys. A: Math. Gen.} {\bf 25} L1;
        1988 {\it Class. Quantum Grav.} {\bf 5} 1259
\bibitem{kleinert} Kleinert H 1989 {\it Gauge Fields in Condensed Matter}
                  vol 1 and 2 (Singapore: World Scientific)
\bibitem{edelen} Kadi\'c A and Edelen D G B 1983 {\it A Gauge Theory of
                Dislocations and Disclinations (Lecture Notes
        in Physics {\bf 174})}, ed H Araki, J Ehlers, K Hepp,
        R Rippenhahn, H A Weidenm\"uller, and J Zittarz
        (Berlin: Springer-Verlag).
\bibitem{lagoud} Edelen D G B and Lagoudas D C 1988 {\it Gauge Theory and
                 Defects in Solids (Mechanics and Physics of Discrete Systems
                 {\bf 1}}) ed G C Sih (Amsterdam: North - Holland)
\bibitem{kondo} Kondo K 1955 Non-Riemannian Geometry of Imperfect
        Crystals from a Macroscopic Viewpoint {\it RAAG Memoirs}
        vol 1 (Gakujutsu Bunken Fukyu-kai, Tokyo)
\bibitem{bilby} Bilby B A, Bullough R and Smith E 1955 {\it Proc. Roy. Soc.
        London A} {\bf 231} 263
\bibitem{jpa91} Osipov V A 1991 {\it J. Phys. A: Math. Gen.} {\bf 24} 3237
\bibitem{jpa93} Osipov V A 1993 {\it J. Phys. A: Math. Gen.} {\bf 26} 1375
\bibitem{pla92} Osipov V A 1992 {\it Phys. Lett. A} {\bf 164} 327
\bibitem{pla94} Osipov V A 1994 {\it Phys. Lett. A} {\bf 193} 97
\bibitem{pa91} Osipov V A 1991 {\it Physica A} {\bf 175} 369
\bibitem{jpc95} Osipov V A and Krasavin S E 1995
                {\it J. Phys.: Cond. Mat.} {\bf 7} L95
\bibitem{prb95} Osipov V A 1995 {\it Phys. Rev. B} {\bf 51} 8614





\end{thebibliography}
\end{document}